# High-fidelity tomography of fluorescent ion qubits under conditions of limited discrimination between bright and dark levels

Yu.I. Bogdanov*, B.I. Bantysh, N.A. Bogdanova, I.A. Dmitriev, V.F. Lukichev
Valiev Institute of Physics and Technology of Russian Academy of Sciences, Moscow, Russia

## ABSTRACT

The present work is devoted to the development of a method for high-precision tomography of ion qubit registers under conditions of limited distinguishability of the states of a logical value 0 and a logical value 1. In the considered ion qubits, the identification of the quantum state is achieved by measuring the fluorescence of the ion by repeated excitation of the cyclic transition, which includes only the lower energy state that sets a logical value 0 and becomes bright, but does not include the upper metastable state that remains dark and sets logical value 1. It is important to note that it is not always possible to achieve low levels of registration errors due to the finite lifetime of excited levels, photon scattering, dark noise, low numerical aperture values, etc. However, even under such conditions, with use of the model of fuzzy quantum measurements, it is possible to provide precise control of quantum states. We show that a model that is characterized by relatively high levels of errors under conditions, where we have a reliable statistical model of their occurrence, is more accurate than the case when the considered errors are small, but we do not have an adequate statistical model for the occurrence of these errors. In the given illustrative examples, we show that the factor of reducing the loss of accuracy with the use of the model of fuzzy measurements can reach values of the order of 1000 or more in comparison with standard measurements.
The obtained results are essential for the development of high-precision methods for controlling the technology of quantum computing on the ion platform.

## 1. INTRODUCTION

To implement precise and repeatable quantum computations, it is necessary to ensure low error levels for quantum state preparation, transformation and readout. Reliable readout of a quantum state is important not only by itself, but also for achieving a correct execution of quantum error correction codes. Trapped ions are promising for quantum information processing. However, the readout errors are quite high.[1-4]

To date, the main mechanism for reading out the state of an ion qubit is the process of fluorescence, which makes it possible to distinguish ion levels by the intensity of their fluorescence after excitation.[5-7] In experimental works this method was used to measure ions states with a hyperfine level structure, such as $^{43}Ca^+$, $^{171}Yb^+$, and others.[8-11] In such experiments, in order to achieve a high readout accuracy (more than 99.9%), the fluorescence time of ions takes hundreds or thousands of microseconds. Long readout times are the main limiting factor in ion-qubit computations. The readout accuracy and duration are influenced by such undesirable factors as off-resonance optical pumping, background and dark counts of photodetectors and their low efficiency. Improving the detection efficiency demonstrates the readout accuracy of 99.85% at $t_{det} = 28.1\,\mu s$ and 99.915% at $t_{det} = 99.8\,\mu s$.[12] The development of readout models that take into account off-resonant optical pumping also helps to reduce the error rate. Such a model made it possible to achieve a readout accuracy of 99.977% without the need to increase the detection efficiency.[13] Meanwhile, recent experimental work shows two-qubit gate times up to $t_{gate} = 1.6\,\mu s$ with fidelity 99.8%, which is at least dozens of times less than readout times shown above.[14]

In this paper, we present a tomography model for ion qubits, which takes into account the presence of errors in the process of state registration. The measurements of the states of ion qubits are based on the readings of the photomultiplier tube of the detector, which records a certain number of counts caused by the emission of photons. The simplest decision-making algorithm is as follows. With a small number of counts below a certain threshold, the result is referred to the state $|0\rangle$ of the qubit. With a larger number of counts above the threshold value, the result is referred to the state $|1\rangle$. Such an algorithm obviously leads to the occurrence of certain readout errors due to the fact that there is a significant probability of attributing the state $|1\rangle$ to the result "0" and vice versa. To eliminate such errors, we use fuzzy measurement operators [15-16] in the process of state reconstruction. Fuzzy operators describe projection to the states $|0\rangle$ and $|1\rangle$ with certain probabilities (for more details, see Sections 3 and 4). The developed model made it possible to significantly increase the accuracy of reconstructing the states of ion qubit registers under conditions of limited distinguishability between ion levels.

---

* bogdanov_yurii@inbox.ru

The work has the following structure. Section 2 provides an overview of the general approach to measuring the states of ion qubits. Section 3 describes a mathematical model of photon statistics for bright and dark levels. Section 4 demonstrates the results of quantum state tomography using the fuzzy measurement approach. Section 5 presents the conclusions of the work.

## 2. ION QUBIT MEASUREMENT THROUGH FLUORECENCE

The state of a qubit $|\psi\rangle = c_0|0\rangle + c_1|1\rangle$ is a superposition of the states of two energy levels of an ion. The state $|0\rangle$ corresponds to the ground state of the ion, which functions as a bright state $|B\rangle$, i.e. fluoresces during measurement. The state $|1\rangle$ corresponds to one of the excited metastable long-lived energy levels, which functions as a dark state $|D\rangle$ due to the absence of fluorescence during its measurement. The task is to measure the considered superposition, i.e. extract information about the corresponding amplitudes of the quantum state through quantum measurements.

The measurement is conducted through a certain short-lived state $|m\rangle$ in the energy structure of the ion.[17] We consider a simple case of a trapped ion exposed to selective laser radiation with a frequency $\omega_{om}$ close to the resonance frequency for the transition $|0\rangle \to |m\rangle$. The population is then moves from the ground level $|0\rangle$ to the excited level $|m\rangle$ (fig.1). The ion from the excited level $|m\rangle$ quickly returns to the ground level $|0\rangle$ emitting a photon due to spontaneous emission.

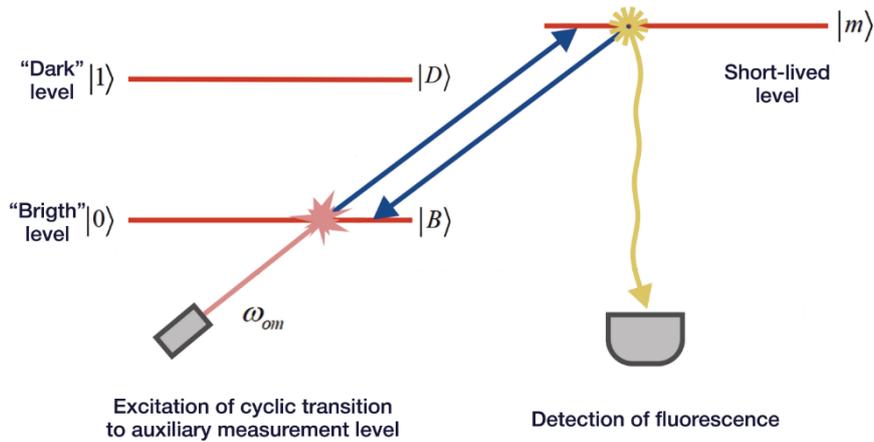

Figure 1. Excitation of a cyclic transition $|0\rangle \to |m\rangle$ by resonant laser radiation.

When this cyclic transition is excited for some time, the ion emits a large number of photons into the solid angle $4\pi$ and some part of the photons are detected by photodetectors. Note that these photons can be separated accurately from the photons of the initial laser radiation propagating in a strictly defined direction. On the other hand, if the ion "hides" at the energy level $|1\rangle$ the fluorescence is not observed, since such radiation is not resonant with the transition $|1\rangle \to |m\rangle$.

# 3. MATHEMATICAL MODEL OF QUBIT FLOURESCENCE

The photon statistics corresponding to the registration of the fluorescence of a bright level over time $t$ is described by the Poisson distribution with the intensity $\lambda_B$:

$$P_B(k) = \frac{(\lambda_B t)^k}{k!} \exp(-\lambda_B t), \qquad (1)$$

where $k$ is the number of registered photons.

In addition to the photons from the bright level fluorescence, the detector also records dark and background counts with the total intensity $\lambda_D$. It is usually several orders of magnitude less than the intensity $\lambda_B$. Dark counts are fundamentally related to the finite temperature of the recording medium and are inherent in any detector. Background counts are associated, in particular, with scattered photons from the primary laser beam. In the ideal case these counts are not present. However, due to the imperfection of the measurement setup, such events are recorded with some small, but finite intensity $\lambda_D$.[18-19]

Let us also take into account the finite lifetime of the metastable level $|1\rangle$. Due to amplitude (energy) relaxation, the level $|1\rangle$ decays exponentially with intensity $\lambda = 1/T_1$, where $T_1$ - is the amplitude relaxation time. If the total observation time is $t$, then there is a nonzero probability of the decay of the level $|1\rangle$ to level $|0\rangle$ at the moment of time $t_1 < t$, after which the population of the level begins to participate in a cyclic transition $|0\rangle \to |m\rangle$ (as a result of this, the state $|1\rangle$ can be mistaken for a state $|0\rangle$).

For further analysis of statistical distributions, let us turn to the mathematical apparatus of generating functions.[20-21] By definition, the generating function for distribution $P(k)$ is

$$G(z) = \sum_{k=0}^{\infty} P(k) z^k, \qquad (2)$$

where $0 \leq z \leq 1$ - is the parameter of the generating function. The generating function of the Poisson distribution is:

$$G(z) = \exp(-\lambda t(z-1)). \qquad (3)$$

Let the upper level $|1\rangle$ decay at the moment of time $t_1$, where $0 \leq t_1 \leq t$. Over time $t - t_1$ the population of the decayed dark level will contribute to the emission of the ion. This case corresponds to the process described by the following generating function:

$$G(z) = \exp(-\lambda_B (t - t_1)(z-1)). \qquad (4)$$

The exponential decay of the upper level with intensity $\lambda = 1/T_1$, which occurs at a moment in time $t_1$, corresponds to the probability density, which is

$$P(t_1) = \lambda \exp(-\lambda t_1). \qquad (5)$$

The generating function of the Poisson distribution (4) must be integrated with the weight given by the density of the exponential distribution (5). As a result, we get

$$G(z|\lambda, \lambda_B, t) = \int_0^t \lambda \exp(-\lambda t_1) \exp(-\lambda_B (t - t_1)(1-z)) dt_1 + \exp(-\lambda t). \qquad (6)$$

Here the second term is responsible for the case when the level is not decayed. Calculating the integral in (6), we get

$$G(z|\lambda, \lambda_B, t) = \exp(-\lambda t) + \frac{\lambda \exp(-\lambda t)\left(\exp\left(t(\lambda - \lambda_B(1-z))\right) - 1\right)}{\lambda - \lambda_B(z-1)}. \tag{7}$$

Note that the generating functions contain complete information about the random variable distribution. In particular, the probability that a discrete random variable takes the value $k$ is proportional to the $k$-th derivative of the generating function at a point $z=0$, and the so-called $m$-th factorial moment is the $m$-th derivative of the generating function at a point $z=1$:

$$P(k) = \frac{1}{k!}\frac{\partial^k G(z)}{\partial z^k}\bigg|_{z=0}, \tag{8}$$

$$M\left[k(k-1)\ldots(k-m+1)\right] = \frac{\partial^m G(z)}{\partial z^m}\bigg|_{z=1}. \tag{9}$$

Here, the symbol $M$ denotes the mathematical expectation operation.

The quantity on the left in (9) is the $m$-th factorial moment. In particular, using the first and second factorial moments, the mean and variance of a random variable can be found:

$$M[k] = G'(1), \tag{10}$$

$$D[k] = G''(1) + G'(1) - G'^2(1). \tag{11}$$

Using formula (8) for the generating function (7) we obtain

$$P(k) = \frac{\lambda t \exp(-\lambda t)(\lambda_B t)^k}{\left((\lambda_B - \lambda)t\right)^{k+1}} \gamma\left((\lambda_B - \lambda)t, k+1\right) + \exp(-\lambda t)\delta_{k0}. \tag{12}$$

Here the second term is nonzero only for $k=0$, which is determined by the Kronecker symbol $\delta_{k0}$. The resulting expression depends on the incomplete gamma function

$$\gamma(x, a) = \frac{1}{\Gamma(a)}\int_0^x t^{a-1}\exp(-t)dt. \tag{13}$$

Using (10) and (7), we obtain the expression for the average number of registered photons:

$$M[k] = \lambda_B t - \frac{\lambda_B}{\lambda}\left(1 - \exp(-\lambda t)\right). \tag{14}$$

Using (9) and (7), we obtain the following expression for the second factorial moment:

$$M[k(k-1)] = (\lambda_B t)^2 \left\{1 - \frac{2}{\lambda t}\left[1 - \frac{1-\exp(-\lambda t)}{\lambda t}\right]\right\}. \tag{15}$$

The variance of the considered distribution is easy to calculate using formulas (11), (14), and (15).

Note that generating function (7) does not take into account the Poisson dark and background noise of the detector with intensity $\lambda_D$. Taking into account that the generating function of the sum of two independent random variables is the product of the generating functions of the distributions of these random variables, we obtain the following result, which takes into account the Poisson noise readings of the detector with the intensity $\lambda_D$:

$$G(z|\lambda, \lambda_B, \lambda_D, t) = \left(\exp(-\lambda t) + \frac{\lambda \exp(-\lambda t)\left(\exp\left(t(\lambda - \lambda_B(1-z))\right) - 1\right)}{\lambda - \lambda_B(z-1)}\right)\exp(-\lambda_D t(1-z)). \tag{16}$$

The desired probability distribution $P_D(k)$ is the convolution of the Poisson distribution and the distribution (12):

$$P_D(k) = \sum_{k_1=0}^{k} \frac{(\lambda_D t)^{k_1}}{k_1!} \exp(-\lambda_D t) P(k-k_1). \qquad (17)$$

The presented calculations are illustrated in Figure 2. This is qualitatively comparable with the results obtained in work[10]. For $P_D(k)$, we see that the most probable number of photons is zero, but nonzero values of the number of photons are also possible in accordance with the theory developed above. For $P_B(k)$, we see that, although the average number of photons is quite large and is equal to 25, a small (or even zero) number of photons can be detected with low probabilities.

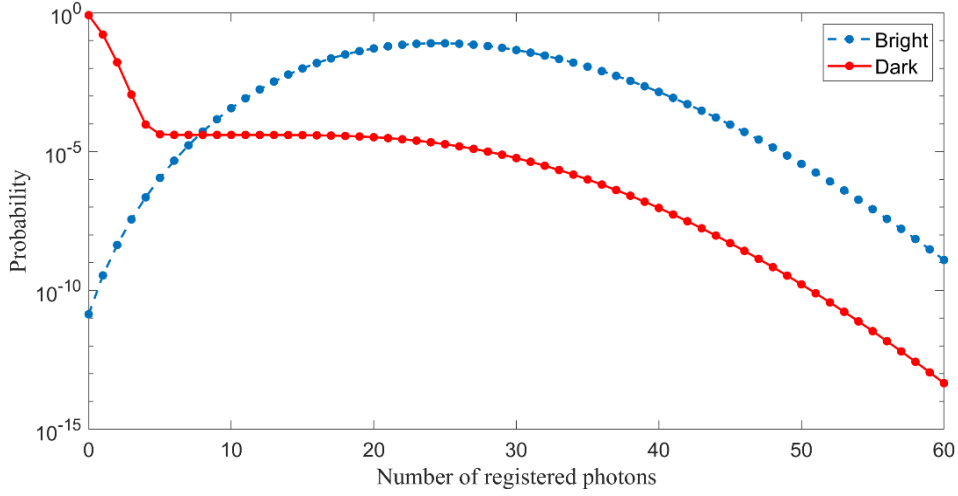

Figure 2. The distributions that appear when registering the state of a fluorescent qubit. The following parameters were selected: $t=1$, $\lambda=0.001$, $\lambda_D=0.2$, $\lambda_B=25$.[10] The red solid curve corresponds to the distribution of the number of photons $P_D(k)$ under conditions when the ion is in the dark state $|D\rangle=|1\rangle$. The blue dashed line corresponds to the distribution of the number of photons $P_B(k)$ under conditions when the ion is in a bright state $|B\rangle=|0\rangle$.

Using the distribution from Figure 2, one must find such a threshold $k_0$, that at $k \geq k_0$ the fluorescence could be considered significant. We take the value where the bright curve becomes higher than the dark one: $k_0=8$ for our case.

The registration process is characterized by the following error rates. The probability of false registration of a dark state $|D\rangle=|1\rangle$ given the ion in the bright state $|B\rangle=|0\rangle$ is

$$P_{error}(1|0) = \sum_{k=0}^{k_0-1} P_B(k). \qquad (18)$$

The probability of false registration of a bright state $|B\rangle=|0\rangle$ given the ion in the dark state $|D\rangle=|1\rangle$:

$$P_{error}(0|1) = \sum_{k=k_0}^{\infty} P_D(k). \qquad (19)$$

For the data presented in Figure 2 we obtain

$$P_{error}(1|0) = 2.292 \cdot 10^{-5}, \; P_{error}(0|1) = 6.878 \cdot 10^{-4}. \qquad (20)$$

The considered example gives rather low error rates. In the next example shown in Figure 3, the corresponding error rates are significantly higher: $k_0 = 1$, $P_{error}(1|0) = 0.0498$, $P_{error}(0|1) = 0.0806$.

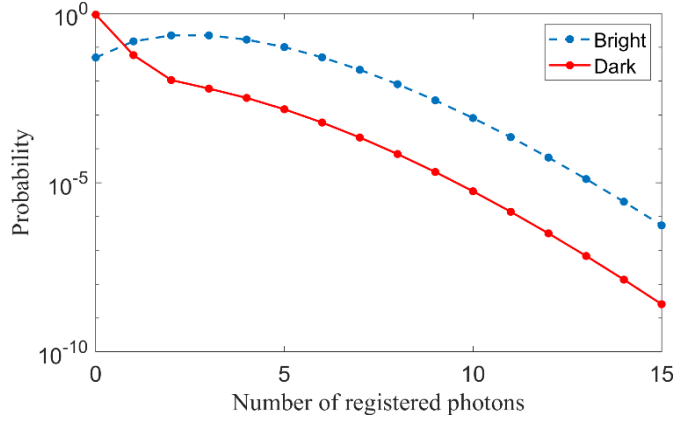

Figure 3. Identification of the state of a fluorescent qubit under conditions of significant registration errors. The following parameters were selected: $t = 1$, $\lambda = 0.05$, $\lambda_D = 0.05$, $\lambda_B = 3$.

The presented errors are quite large. However, as we will see below, the fuzzy measurements model could still provide a high tomography accuracy.

## 4. FUZZY MEASUREMENTS MODEL

Denote measurement error probability as $p_{10} = P_{error}(1|0)$ and $p_{01} = P_{error}(0|1)$. We assume that the initial state of the ion qubit is a superposition of states $|0\rangle$ and $|1\rangle$: $|\psi\rangle = c_0|0\rangle + c_1|1\rangle$. The total probability to get the measurement result "0" is $P_0 = (1-p_{10})|c_0|^2 + p_{01}|c_1|^2$. Here $|c_0|^2$ is the probability to get "0" in the case of no errors. The negative term $-p_{10}|c_0|^2$ characterizes the false registration of the state "1" when the true state is "0". The positive term $p_{01}|c_1|^2$ describes the false registration of the state "0" when the true state is "1". The measurement operator corresponding to the considered probability $P_0$ is $\Lambda_0 = (1-p_{10})|0\rangle\langle 0| + p_{01}|1\rangle\langle 1|$. Similarly, for the result "1" we have $P_1 = (1-p_{01})|c_1|^2 + p_{10}|c_0|^2$ and $\Lambda_1 = p_{10}|0\rangle\langle 0| + (1-p_{01})|1\rangle\langle 1|$.

These measurement operators form POVM measurement: $\Lambda_0 \geq 0$, $\Lambda_1 \geq 0$, $\Lambda_0 + \Lambda_1 = I$.

Obtaining complete information about the quantum state requires measurements in different bases. For this, the state is subjected to some unitary transformation $U$ before the readout. Then, the new measurement operators are

$$\Lambda_0^U = U^+ \Lambda_0 U, \ \Lambda_1^U = U^+ \Lambda_1 U. \tag{21}$$

The introduced fuzzy measurement operators $\Lambda_0^U$ and $\Lambda_1^U$ also define the unity decomposition $\Lambda_0^U + \Lambda_1^U = I$ due to $U^+ U = I$.

In the standard model, each measurement outcome is deterministically related to either projector $|0\rangle\langle 0|$ or projector $|1\rangle\langle 1|$, which inevitably leads to systematic errors in the state reconstruction. As a result, the quantum state tomography model becomes inadequate. On the other hand, the fuzzy measurement operators $\Lambda_0$ and $\Lambda_1$ relate each measurement outcome to $|0\rangle\langle 0|$ and $|1\rangle\langle 1|$ with some probabilities. These probabilities are estimated from the preliminary measurements of bright and dark states fluorescence, as described in Section 3. This approach allows one to correctly describe the measurement statistics even in the presence of readout errors. As a result, the tomography model becomes adequate.

In the case of measuring several qubits, we assume that each qubit is measured independently, and the readout errors for them are identical. However, this limitation is not essential, and the model can be constructed even if these conditions are violated.

Figure 4 shows the tomography accuracy for different measurements models. We consider tomography of 200 two-qubit pure states, taken randomly according to the Haar measure.[22] The quantum states reconstruction was done using the root approach and maximum likelihood method.[23] Tomography protocol is based on Pauli matrices bases.[24] Sample size in each experiment is $N = 10^6$, probabilities of false registration are $p_{01} = p_{10} = 0.1$. Theoretical infidelity distribution was obtained with complete information matrix.[25] In particular, the curve in Figure 4b is the average over the curves for each state separately. We used fuzzy operators to generate statistical data for imperfect measurements. The reconstruction was carried out using two different measurement models: one takes into account readout errors (fuzzy model), and the other does not (standard model). The sample infidelity for the fuzzy measurements model is $\langle 1-F \rangle_{SM} / \langle 1-F \rangle_{FM} = 1129$ times lower than for the standard model of ideal measurements.

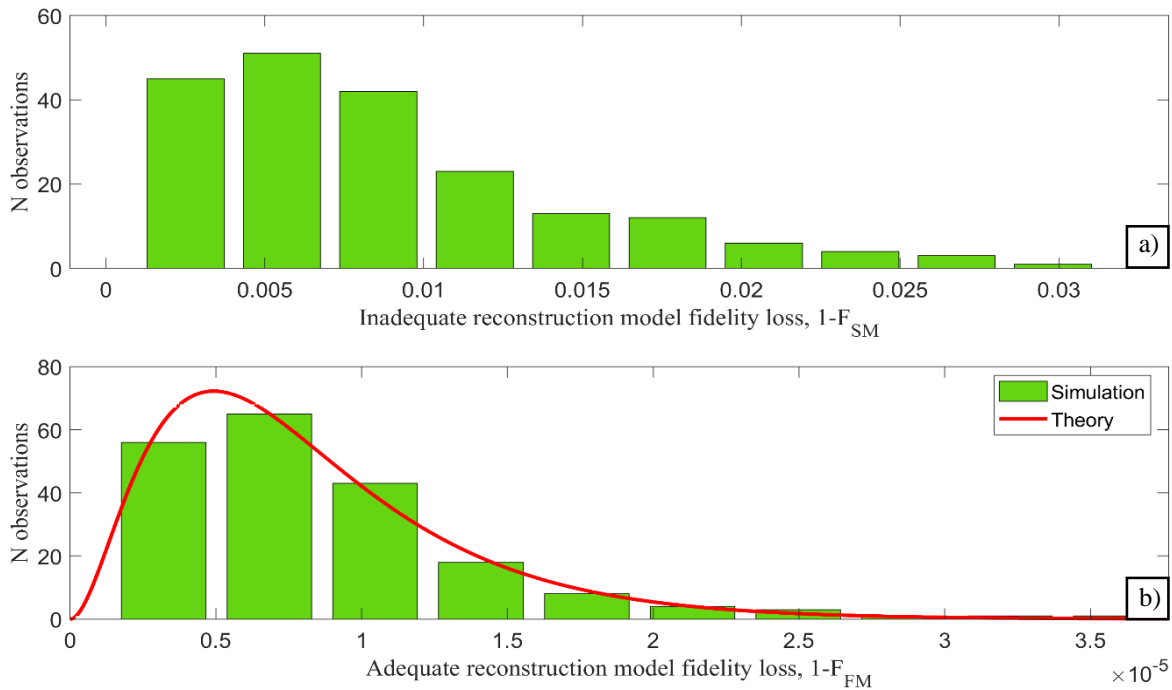

Figure 4. Comparison of the loss of fidelity of tomography when using two different reconstruction models. a) - inadequate (standard) model, b) - model of fuzzy measurements.

It is not necessary to wait for an obvious distinction between signals from bright and dark levels during readout process. Therefore, it is possible to reduce the detection time without a significant loss in the reconstruction fidelities using the fuzzy measurement model.

Note that this kind of fuzzy measurements model is applicable not only for the quantum tomography, but for any experiments that consider collecting the statistical data over a quantum state.

## 5. CONCLUSIONS

The presence of readout errors significantly limits the duration and accuracy of the trapped ion qubit readout. In this work, we have developed the fuzzy measurements model. It allows one to drastically reduce reconstruction fidelity losses under presence of significant levels of readout errors. The statistical model for detecting ion qubit fluorescence has also been developed. The model makes it possible to estimate the readout errors using values of fluorescence intensity, decay time of the dark level, dark counts and measurement time. Thus, the development and application of a fuzzy measurement model makes a significant contribution to improving the accuracy of reconstructing the quantum states of ion qubits under conditions of limited distinguishability of bright and dark states.

## ACKNOWLEDGMENTS

This work was supported the Program of activities of the leading research center "Development of an experimental prototype of a hardware and software complex for the technology of quantum computing based on ions" (Agreement No. 014/20), and by the Foundation for the Advancement of Theoretical Physics and Mathematics BASIS (project no. 20-1-1-34-1).

## REFERENCES

[1] Bruzewicz, C. D., Chiaverini, J., McConnell, R., & Sage, J. M., "Trapped-ion quantum computing: Progress and challenges," *Applied Physics Reviews*, **6**(2), 021314 (2019).
[2] Wineland, D. J., Monroe, C., Itano, W. M., Leibfried, D., King, B. E., & Meekhof, D. M., "Experimental issues in coherent quantum-state manipulation of trapped atomic ions," *Journal of research of the National Institute of Standards and Technology*, **103**(3), 259 (1998).
[3] Blatt, R., & Wineland, D., "Entangled states of trapped atomic ions," *Nature*, **453**(7198), 1008-1015 (2008).
[4] Borisenko, A., Zalivako, I., Semerikov, I., Aksenov, M., Khabarova, K., & Kolachevsky, N., "Motional states of laser cooled Yb ions in an optimized radiofrequency trap," *Laser Physics*, **29**(9), 095201 (2019).
[5] Nagourney, W., Sandberg, J., & Dehmelt, H., "Shelved optical electron amplifier: Observation of quantum jumps," *Physical Review Letters*, **56**(26), 2797 (1986).
[6] Bergquist, J. C., Hulet, R. G., Itano, W. M., & Wineland, D. J., "Observation of quantum jumps in a single atom," *Physical review letters*, **57**(14), 1699 (1986).
[7] Sauter, T., Neuhauser, W., Blatt, R., & Toschek, P. E., "Observation of quantum jumps," *Physical review letters*, **57**(14), 1696 (1986).
[8] Itano, W. M., Bergquist, J. C., Hulet, R. G., & Wineland, D. J., "Precise optical spectroscopy with ion traps," *Physica Scripta*, **1988**(T22), 79 (1988).
[9] Brown, K. R., Wilson, A. C., Colombe, Y., Ospelkaus, C., Meier, A. M., Knill, E., ... & Wineland, D. J., "Single-qubit-gate error below $10^{-4}$ in a trapped ion," *Physical Review A*, **84**(3), 030303 (2011).
[10] Myerson, A. H., Szwer, D. J., Webster, S. C., Allcock, D. T. C., Curtis, M. J., Imreh, G., ... & Lucas, D. M., "High-fidelity readout of trapped-ion qubits," *Physical Review Letters*, **100**(20), 200502 (2008).
[11] Burrell, A. H., *High fidelity readout of trapped ion qubits,* PhD thesis, Oxford University, UK (2010).
[12] Noek, R., Vrijsen, G., Gaultney, D., Mount, E., Kim, T., Maunz, P., & Kim, J., "High speed, high fidelity detection of an atomic hyperfine qubit," *Optics letters*, **38**(22), 4735-4738 (2013).
[13] Wölk, S., Piltz, C., Sriarunothai, T., & Wunderlich, C., "State selective detection of hyperfine qubits," *Journal of Physics B: Atomic, Molecular and Optical Physics*, **48**(7), 075101 (2015).
[14] Schäfer, V. M., Ballance, C. J., Thirumalai, K., Stephenson, L. J., Ballance, T. G., Steane, A. M., & Lucas, D. M., "Fast quantum logic gates with trapped-ion qubits," *Nature*, **555**(7694), 75-78 (2018).
[15] Bogdanov, Y. I., Bantysh, B. I., Bogdanova, N. A., Kvasnyy, A. B., & Lukichev, V. F., ''Quantum states tomography with noisy measurement channels'', in [International Conference on Micro-and Nano-Electronics 2016], Proc. SPIE **10224**, 102242O (2015).
[16] Bantysh, B. I., Bogdanov, Y. I., Bogdanova, N. A., & Kuznetsov, Y. A., "Precise tomography of optical polarisation qubits under conditions of chromatic aberration of quantum transformations," *Laser Physics Letters*, **17**(3), 035205 (2020).
[17] Low, P. J., White, B. M., Cox, A. A., Day, M. L., & Senko, C., "Practical trapped-ion protocols for universal qudit-based quantum computing," *Physical Review Research*, **2**(3) 033128, (2020).
[18] Hadfield, R. H., "Single-photon detectors for optical quantum information applications," *Nature photonics*, **3**(12), 696-705 (2009).

[19] Todaro, S. L., Verma, V. B., McCormick, K. C., Allcock, D. T. C., Mirin, R. P., Wineland, D. J., Slichter, D. H., "State readout of a trapped ion qubit using a trap-integrated superconducting photon detector," *Physical Review Letters*, **126**(1), 010501 (2021).

[20] Feller, W., [An introduction to probability theory and its applications, vol 2], John Wiley & Sons (2008).

[21] Bogdanov, Y. I., Bogdanova, N. A., Katamadze, K. G., Avosopyants, G. V., & Lukichev, V. F., "Study of photon statistics using a compound Poisson distribution and quadrature measurements," *Optoelectronics, Instrumentation and Data Processing*, **52**(5), 475-485 (2016).

[22] Hayden, P., Leung, D., Shor, P. W., & Winter, A., "Randomizing quantum states: Constructions and applications," *Communications in Mathematical Physics*, **250**(2), 371-391 (2004).

[23] Bogdanov, Y. I., "Fundamental notions of classical and quantum statistics: A root approach," *Optics and spectroscopy*, **96**(5), 668-678 (2004).

[24] Adamson, R. B. A., & Steinberg, A. M., "Improving quantum state estimation with mutually unbiased bases," *Physical review letters*, **105**(3), 030406 (2010).

[25] Bogdanov, Y. I., Chekhova, M. V., Krivitsky, L. A., Kulik, S. P., Penin, A. N., Zhukov, Tey, M. K., "Statistical reconstruction of qutrits," *Physical Review A*, **70**(4), 042303 (2004).